\documentclass[aps,pre,preprint,superscriptaddress]{revtex4-2}

\usepackage{physics}
\usepackage{amssymb}
\usepackage[dvipdfmx]{graphicx}
\usepackage{here}
\usepackage{MnSymbol}

\begin{document}


\title{Convergence study of multi-field singular value decomposition for turbulence fields}


\author{Go Yatomi}
\email{yatomi.go@nifs.ac.jp}
\affiliation{SOKENDAI (The Graduate University for Advanced Science), Gifu 509-5292, Japan}
\author{Motoki Nakata}
\affiliation{Faculty of Arts and Sciences, Komazawa University, Tokyo 154-8525, Japan \\
RIKEN Interdisciplinary Theoretical and Mathematical Sciences Program (iTHEMS), Saitama 351-0198, Japan \\
PRESTO, Japan Science and Technology Agency, Saitama 332-0012, Japan}


\date{\today}

\begin{abstract}
  Convergence of a matrix decomposition technique, the multi-field singular value decomposition (MFSVD) which efficiently analyzes nonlinear correlations by simultaneously decomposing multiple fields, is investigated. Toward applications in turbulence studies, we demonstrate that SVD for an artificial matrix with multi-scale structures reproduces the power-law-like distribution in the singular value spectrum with several orthogonal modes. Then, MFSVD is applied to practical turbulence field data produced by numerical simulations. It is clarified that relative errors in the reproduction of quadratic nonlinear quantities in multi-field turbulence converge remarkably faster than the single-field case, which requires thousands of modes to converge.
\end{abstract}


\maketitle

\section{Introduction}
In recent decades, the size of data treated in natural science has kept increasing. Due to the development of high-performance computers and ameliorated algorithms, high-resolution and long-term simulations are realized. Since the data obtained from these simulations become so huge and complicated, it is no longer straightforward to fully analyze and understand the features. More elaborate approaches that can extract the essential characteristics from the huge data are required. \par
In physics, astronomy, and engineering, turbulence in fluids and plasmas is one of the long-standing problems treating high-resolution data in numerical simulations, laboratory experiments, and observations. Turbulence is composed of eddies with various spatial scales, and the power-law-like spectrum called the Kolmogorov's law \cite{Davidson2015} often appears in the Fourier wavenumber space. Moreover, turbulence often consists of multiple filed quantities such as density field, temperature fields, velocity field, and so on. High-resolution and multi-component/multi-field simulations including the wide range of eddy scales are thus indispensable for the studies of fluid and plasma dynamics \cite{Gotoh2002,Maeyama2022}. \par
To extract the essential information from these huge data in turbulence, data-driven methods have been developed. Some powerful data-driven analyses are based on matrix decompositions such as the singular value decomposition \cite{Lim2004,Sasaki2021} and the dynamical mode decomposition \cite{Pan2011,Sasaki2019} for general matrices. While these analyses can extract characteristic modes of the original flow field, they usually treat only one field quantity, and the correlations between multiple fields are ignored. Then, we recently proposed a novel method referred to as the ''multi-field singular value decomposition'' (MFSVD) \cite{Yatomi2023}, which decomposes the multiple field quantities simultaneously.  \par
In this work, we present further investigations of the numerical convergence in the SVD and MFSVD for the turbulent fields and their nonlinear correlations. The SVD has conventionally been used for the low-rank approximation of laminar or weakly turbulent flows such as the K\'{a}rm\'{a}n vortex streets that appear in the flow behind a cylindrical object. However, according to the author's knowledge, the fundamental convergence characteristics of the matrix decompositions for the strongly turbulent state with a significant multi-scale nature are still unclear. Here, we examine the mode-number dependence of reproduction accuracy for the single turbulent field and the quadratic correlations by means of both an artificial multi-scale field and realistic turbulence fields. As will be shown later, it is confirmed that the nonlinear correlations of fields can be expressed smaller number of modes than that in the original single field, indicating the capability of low-rank approximation of strongly turbulent states using the MFSVD.

\section{Multi-field singular value decomposition}

\begin{figure}[tb]
  \centering
  \includegraphics[width=14cm]{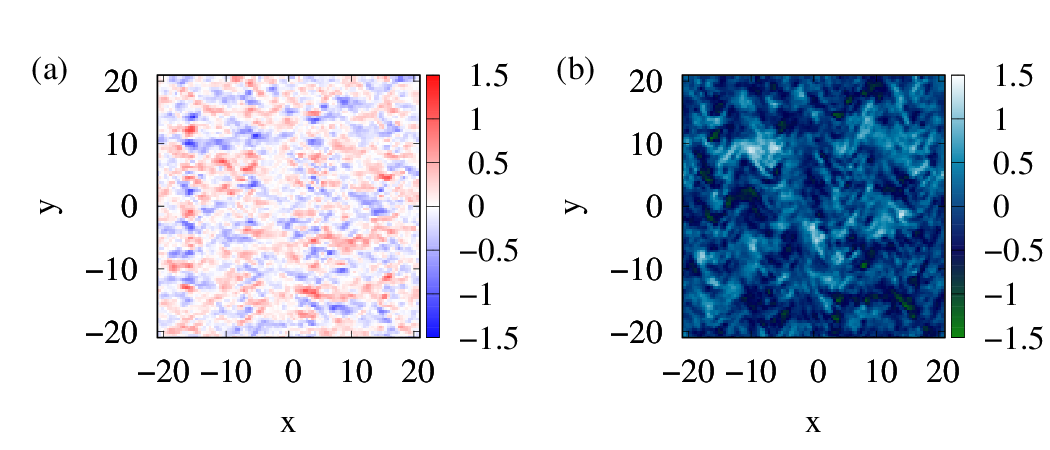}
  \caption{Snapshots of the turbulent fields obtained from a numerical simulation of a fluid model. (a) shows the velocity component $v$ and (b) shows the particle density $n$.}
  \label{fig:fields}
\end{figure}

The procedure of a matrix decomposition for a general turbulence system with multiple field quantities is explained as follows, where the singular value decomposition is utilized as a fundamental technique \cite{Yatomi2023}. We consider $N_F$ fields of time-dependent physical quantities $g_l(x,y,t)$ ($l=1,\cdots,N_F$) in two-dimensional space for simplicity. The extension to the higher-dimensional cases is straightforward. Fig. \ref{fig:fields} shows examples of snapshots of field quantities, i.e., the velocity component $v$ and the particle density $n$ in a turbulence system \cite{Hasegawa1983,Numata2007}. These two-dimensional field data are obtained by the numerical simulations of coupled nonlinear field equations under the periodic boundary conditions with respect to x and y. \par
Let $g_l(x,y,t)$ be a general two-dimensional field for the $l$-th component. These fields are discretized in both temporal and spatial variables, i.e., $g_l(x_i,y_j,t_k)$ ($i=1,\cdots N_x$, $j=1,\cdots N_y$, $k=1,\cdots N_t$), where $N_x$ and $N_y$ are the grid number in the $x$, $y$, and the time direction, respectively. Then, each field at each time is rearranged to a time-dependent $(N_xN_y)$-dimension vector: 
\begin{align}
  \vb{g}_l(t_k) = \left(g_l(x_1,y_1,t_k),\cdots,g_l(x_{N_x},y_1,t_k),\cdots, \right. \nonumber \\
  \left. g_l(x_1,y_{N_y},t_k),\cdots,g_l(x_{N_x},y_{N_y},t_k)\right)^\mathrm{T}.
\end{align}
These vectors are aligned to a $(N_xN_y)\times(N_tN_F)$ matrix $G$ such as
\begin{align}
  G = (\vb{g}_1(t_1),\cdots,\vb{g}_{N_F}(t_1),\cdots,\vb{g}_1(t_{N_t}),\cdots,\vb{g}_{N_F}(t_{N_t})).
\end{align}
By letting $N:=\min(N_xN_y,N_tN_F$), the matrix $G$ is decomposed using three matrices:
\begin{align}
  G = U\Sigma V^\mathrm{T},
\end{align}
where $U$ is a $(N_xN_y)\times N$ matrix of the left eigenvectors of the covariance matrix $G^\mathrm{T}G$, $V$ is a $(N_tN_F)\times N$ matrix of the right eigenvectors of $G^\mathrm{T}G$, and $\Sigma$ is a $N\times N$ diagonal matrix that contains the ''singular value'' $s_i$, which is the square root of the $i$-th eigenvalue of $G^\mathrm{T}G$, in the $i$-th diagonal component. Note that we assume here $G$ to be real-valued, the same arguments hold for complex-valued matrices as well. From the way the matrix $G$ is constructed, each column of $U$ corresponds to the spatial structure of each mode and it is denoted as $\psi_i(x,y)$ so that $\psi_i(x_j,y_k)=U_{j+N_y(k-1),i}$. Similarly, each column of $V$ corresponds to the temporal evolution of each mode and it is denoted as $h_i^{(l)}(t)$ so that $h_i^{(l)}(t_m)=V_{l+N_t(m-1),i}$. Then, in the continuous limit, the field quantity $g_l(x,y,t)$ is expressed by a series of the orthogonal functions characterized by the singular value:
\begin{align}
  g_l(x,y,t) = \sum_{i=1}^N s_i h_i^{(l)}(t)\psi_i(x,y).
\end{align} 
We emphasize that only the temporal evolution coefficient $h_i^{(l)}(t)$ depends on the field index and the basis of the spatial structure $\psi_i(x,y)$ is common to all SVD modes. Therefore, one can treat all fields equally, and phase relations between fields are straightforwardly analyzed as explained below. \par
Note that the spatial structures $\psi_i$ satisfies the orthonormal condition, i.e., 
\begin{align}
  \int\dd\vb{x}\,\psi_i\psi_j:= \sum_{a=1}^{N_x}\sum_{b=1}^{N_y}\psi_i(x_a,y_b)\psi_j(x_a,y_b) = \delta_{ij},
\end{align}
because $U$ is the matrix of eigenvectors of the real-symmetric matrix $G^\mathrm{T}G$. Thus, the second-order quantity of the fields is also decomposed by SVD modes as
\begin{align}
  q &:= \int\dd\vb{x}\,g_1(x,y,t)g_2(x,y,t) \nonumber \\
    &= \sum_{i=1}^N s_i^2h_i^{(1)}(t)h_i^{(2)}(t) =: \sum_{i=1}^Nq_i.
  \label{eq:SVD_of_NL}
\end{align} \par
Note that the field quantity we decompose can also be regarded as a 3rd or 4th order tensorial data with three or four modes, i.e., the spatial directions in x and/or y, time, and field type. Thus, we can apply  tensor decomposition techniques such as the CP decomposition and Tucker decomposition \cite{Coley2022,Alfaro202,Zhang2022}, but some care about the non-uniqueness of decomposition and the large numerical costs must be taken in applying them. On the other hand, the present MFSVD intentionally performs the matrix decomposition by contracting the multi-component mode and arranging them in columns to obtain a uniquely determined singular value distribution, from the physical motivation to treat the quadratic interactions of fields. Indeed, the nonlinear correlations, which are quite important in the physics of turbulence, are also simply decomposed as shown in Eq. \eqref{eq:SVD_of_NL}. This is the novel feature of the present extension of the SVD to the multi-component fields.

\section{Matrix decomposition for multi-scale structures}
Turbulence fields generally exhibit several complexities in the spatio-temporal dynamics, e.g., multi-scale nature, inhomogeneity, and non-locality. It seems the ensembles of stochastic patterns, but a statistical power law known as Kolmogorov's law exists in the turbulence, where the energy of an eddy with the size of $l$ and the wavenumber of $k =1/l$ scales as $E_k \propto k^{-5/3}$ \cite{Davidson2015}. In order to clarify the fundamental characteristics of the matrix decomposition for the turbulent fields, we examine the singular value distribution and the mode structure in the SVD for a 100-dimensional square matrix, where the multi-scale property is artificially imposed to each entry in the matrix. \par
\begin{figure}[tb]
  \centering
  \includegraphics[width=14cm]{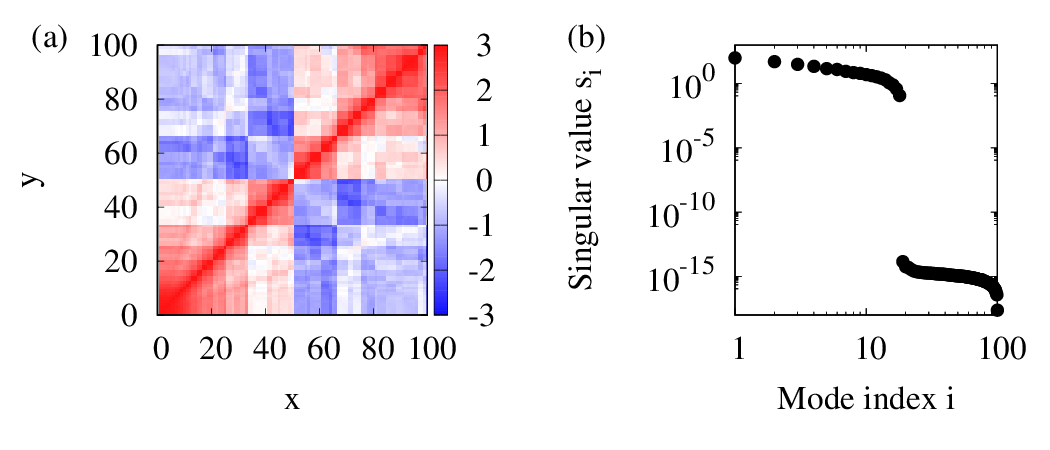}
  \caption{Application of the SVD to the multi-scale checkerboard matrix $F$. (a) shows the input matrix and (b) shows the singular value spectrum.}
  \label{fig:svd_chb}
\end{figure}

As an imitation of a single turbulent field, a multi-scale ''checkerboard'' matrix is considered. Let $p$ be an integer representing the size of the checkerboard and blocks of $p$-dimensional square matrices are aligned in a $100$-dimensional square matrix. The value for each block is assigned $+1$ or $-1$ so that neighboring blocks have opposite signs, denoting this matrix by $F_p$. The matrix $F$ is defined as a superposition of $F_p$ with a weight $w_p$:
\begin{align}
  F = \sum_{p\in P} w_p F_p.
\end{align}
Here, the size of the checkerboard $p$ is set to $p\in P=\{1,2,3,4,5,6,7,8,9,10,11,12,14,16,20,25,33,50\}$ and the weight is to be proportional to $p^{-5/3}$ in order to imitate the turbulence with wavenumbers $k=1/p$ whose amplitude follows the Kolmogorov's law. The contour of the input matrix $F$ is shown in Fig. \ref{fig:svd_chb}(a), where $w_{50}=1$. \par
\begin{figure}[tb]
  \centering
  \includegraphics[width=14cm]{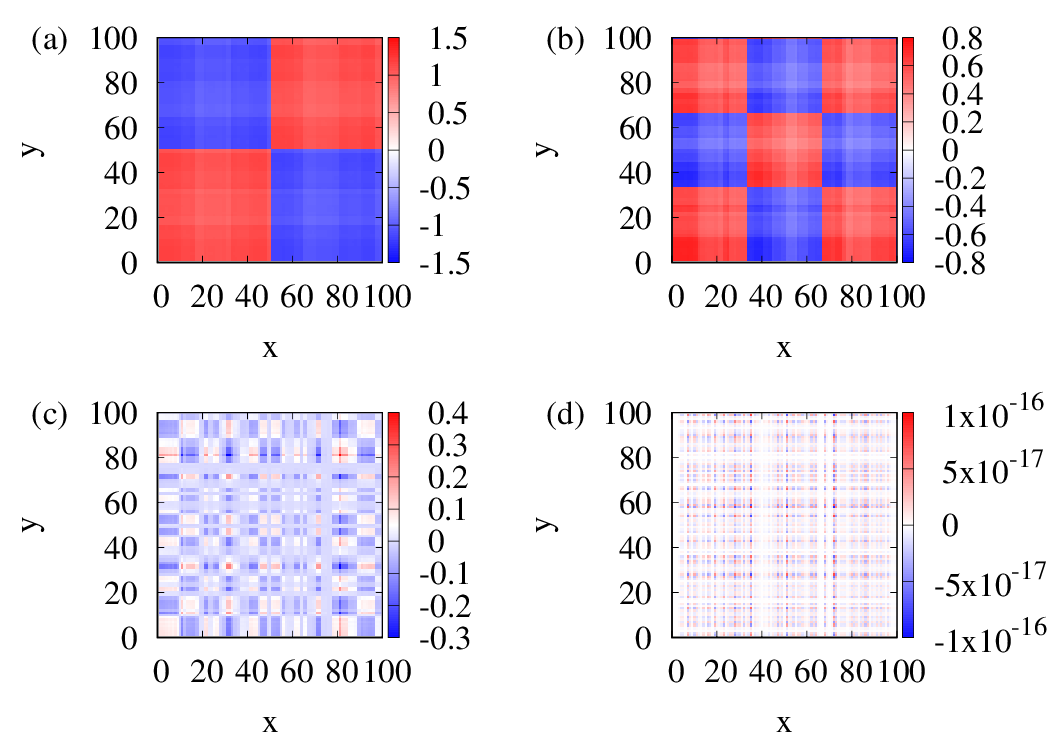}
  \caption{The contours of the eigenmode of the input matrix $F$. Modes 1, 2, 10, and 50 are shown in panels (a), (b), (c), and (d), respectively.}
  \label{fig:modes}
\end{figure}

Fig. \ref{fig:svd_chb}(b) shows the spectrum of singular values when the SVD is applied to $F$. The singular values are mostly dominated by the first $18$ modes which is consistent with the number of imposed checkerboards $F_p$, while the singular values decrease by more than $10$ orders of magnitude for the later modes associated with the noisy patterns. This is because each scale of the checkerboard $F_p$ is a nearly orthogonal matrix, and consequently the singular value decomposed basis is close to the original $F_p$. Through these experiments, we demonstrated that the SVD-based matrix decomposition can successfully decompose the turbulence field with a power-law-like scaling, where the multi-scale nature is projected into the singular value spectrum and the associated mode patterns. This is also confirmed from the contour plots of the SVD mode structures shown in Fig. \ref{fig:modes}.

\section{Convergence for multiple fields and the nonlinear correlations}
In order to examine the capability of the low-rank approximation or dimensionality reduction for realistic turbulence fields, we investigate the convergence with respect to the reproduction accuracy of MFSVD. Here, two-dimensional turbulence with multiple field quantities in magnetized plasmas is considered, where the field data is produced by the numerical simulations of the nonlinear dynamical equation called Hasegawa-Wakatani equation \cite{Yatomi2023,Hasegawa1983,Numata2007}.  \par
One important nonlinear correlation quantity in fluid and plasma turbulence is the quadratic form of the field quantities such as the so-called particle flux $\Gamma$, expressed by the product of the particle density $n$ and the velocity component $v$ as
\begin{align}
  \Gamma := \int\dd\vb{x}\,v(x,y,t)n(x,y,t).
\end{align}
Then, the MFSVD is applied to the dataset of the fields $n$ and $v$, which are obtained from the numerical simulation: $v(x,y,t)=\sum_{i=1}^Ns_ih_i^{(v)}(t)\psi_i(x,y)$ and $n(x,y,t)=\sum_{i=1}^Ns_ih_i^{(n)}(t)\psi_i(x,y,t)$. Here, the total number of the SVD modes $N$ is set to $10^4$. Since the particle flux $\Gamma$ is a quadratic quantity of fields, it is also simply decomposed into the SVD modes as shown in Eq. \eqref{eq:SVD_of_NL}:
\begin{align}
  \Gamma = \sum_i\Gamma_{n,i}=\sum_is_i^2h_i^{(v)}h_i^{(n)}.
  \label{eq:SVD_gamma}
\end{align}
For the original field quantity $v$ and the particle flux $\Gamma$, we define a relative deviation between the exact value and the value with truncation at the mode $k$: 
\begin{align}
  \Delta_k^{(v)} &:= \int\dd t\dd\vb{x}\left| \left(v-\sum_{i=1}^ks_ih_i^{(v)}\psi_i\right)\middle/v \right|, \\
  \Delta_k^{(\Gamma)} &:= \int\dd t\left| \left(\Gamma - \sum_{i=1}^k\Gamma_{n,i}\right)\middle/\Gamma\right|.
\end{align} \par
\begin{figure}[tb]
  \centering
  \includegraphics[width=14cm]{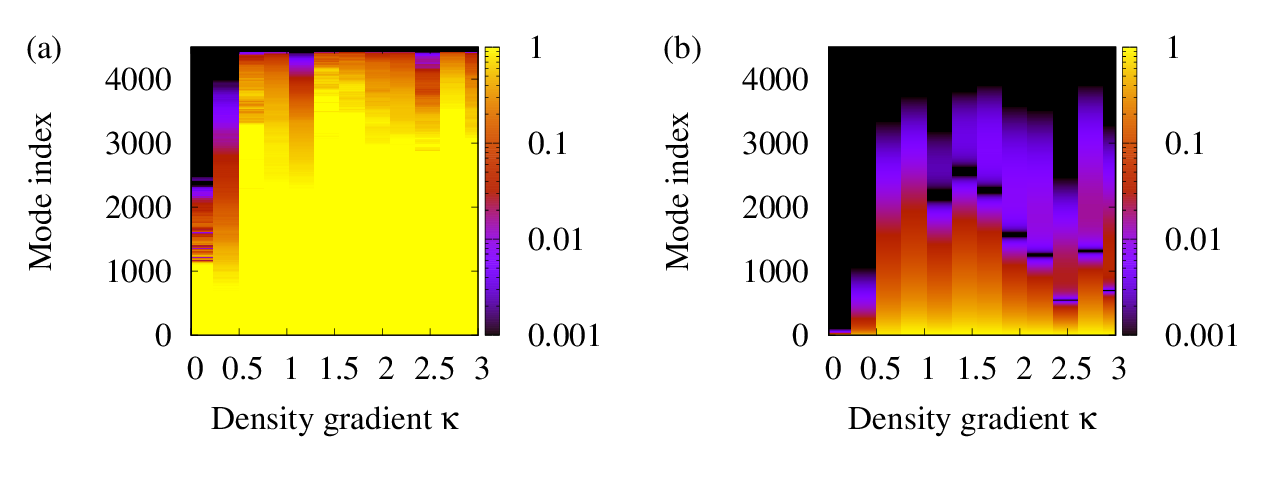}
  \caption{The $\kappa$ dependence of the convergence of the (a) the field value $v$ and (b) the nonlinear particle flux $\Gamma$.}
  \label{fig:converge}
\end{figure}

In the Hasegawa-Wakatani equation, the turbulence intensity is controlled by a constant physical parameter $\kappa$ which denotes the strength of the background inhomogeneity. Then, we investigate the $\kappa$ dependence of the convergence to look into the difference between the nearly-laminar state for $\kappa<1$ and the strongly turbulent state for $\kappa<1$. Fig. \ref{fig:converge}(a) shows the $\kappa$ dependence of the convergence of the original field $v$. In most $\kappa$ regions, the relative deviation $\Delta_k^{(v)}$ is reduced by only about two orders of magnitude, even at mode numbers of about 4000. This is because the vortices with a wide range of spatial scales are present in the original turbulent field, and these must be included sufficiently in the SVD modes to recover the field values. In the case of the nearly-laminar state for $\kappa<1$, the convergence is relatively fast compared to large $\kappa$ cases. It is noted that thousands of modes are still needed to express the original field value with sufficient accuracy. The results are qualitatively consistent with the multi-scale structures presented in Sec. 3. \par
On the other hand, the relative deviation of the particle flux $\Delta_k^{(\Gamma)}$ is concentrated in earlier mode numbers and decreases more rapidly as shown in Fig. \ref{fig:converge}(b). In the expression of the particle flux, Eq. \eqref{eq:SVD_gamma}, the factor of the phase correlation $h_i^{(v)}h_i^{(n)}$ appears. Therefore, coherent modes with significant residual values after the time integration contribute to the quadratic quantity $\Gamma$. These results can be rephrased as follows. Even though a number of SVD modes are necessary to reproduce the single turbulent field with reasonable accuracy, the nonlinear correlation of the two fields remarkably reduces the number of modes required for reproduction, where the random components in the phase difference do not much contribute to the nonlinear correlation. The present convergence study demonstrates that the MFSVD can express the nonlinear quantities, which are important in the physics of turbulent fluids, with a reduced number of modes.

\section{Summary}
In this work, the convergence characteristics of the multi-field singular value decomposition (MFSVD) are investigated. The application of SVD to artificial matrix data that mimic turbulent fields confirms that SVD decomposes multi-scale turbulent fields into individual spatial scales with orthogonal structures. \par
The MFSVD is applied to the realistic turbulence field obtained by numerical simulations to investigate the convergence of the relative deviation from the original turbulence field and the nonlinear quantities. Convergence for the single turbulence field takes several thousand modes, reflecting the fact that eddies with a huge number of length scales are superimposed in turbulent fields. On the other hand, the particle flux, which is a representative example of a nonlinear quantity, showed faster convergence than the single turbulence field. This is because the MFSVD can efficiently extract the coherent modes that contribute to the average quantity by including the phase correlations among multiple fields in the decomposition. \par
The ability of MFSVD to recover nonlinear quantities with a reduced number of modes indicates the capability of using the MFSVD to construct a reduced model of multi-field turbulence. Using the basis functions $\psi_i(x,y)$ obtained by the MFSVD to derive time evolution equations for nonlinear quantities, it is expected to capture the characteristic nonlinear phenomena of turbulent fields with less computational cost. Another future work is to consider information-theoretic entropy by focusing on the orthonormality of the basis functions.

\section*{Acknowledgments}
Numerical computations are performed on the NIFS Plasma Simulator. This work was supported in part by JST, the establishment of university fellowships towards the creation of science technology innovation, Grant Number JPMJFS2136, in part by JST, PRESTO Grant Number JPMJPR21O7, and in part by the NIFS collaborative Research programs (NIFS23KIST039, NIFS23KIST044).


\begin{thebibliography}{99}
\bibitem{Davidson2015}
P.~Davidson, Turbulence: An Introduction for Scientists and Engineers, 2nd edn., Oxford University Press, Oxford, 2015.

\bibitem{Gotoh2002}
T.~Gotoh, D.~Fukayama, and T.~Nakano, 
Velocity field statistics in homogeneous steady turbulence obtained using a high-resolution direct numerical simulation, Phys. Fluids, {\bf 14} (2002), 1065.

\bibitem{Maeyama2022}
S.~Maeyama et al., 
Multi-scale turbulence simulation suggesting improvement of electron heated plasma confinement, Nat. Commun., {\bf 13} (2022), 3166.

\bibitem{Lim2004}
J.~Lim and J.~Kim, 
A singular value analysis of boundary layer control, Phys. Fluids, {\bf 16} (2004), 1980.

\bibitem{Sasaki2021}
M.~Sasaki et al., 
Evaluation of abrupt energy transfer among turbulent plasma structures using singular value decomposition, Plasma Phys. Cntrol. Fusion, {\bf 63} (2021), 025004.

\bibitem{Pan2011}
C.~Pan, D.~Yu, and J.~Wang, 
Dynamical mode decomposition of Gurney flap wake flow, Theor. Appl. Mech. Lett., {\bf 1} (2011), 012002.

\bibitem{Sasaki2019}
M.~Sasaki et al., 
Using dynamical mode decomposition to extract the limit cycle dynamics of modulated turbulence in a plasma simulation, Plasma Phys. Control. Fusion, {\bf 62} (2019), 112001.

\bibitem{Yatomi2023}
G.~Yatomi, M.~Nakata, and M.~Sasaki, 
Data-driven modal analysis of nonlinear quantities in turbulent plasmas using multi-field singular value decomposition,
Plasma Phys. Control. Fusion., {\bf 65} (2023), 
095014.

\bibitem{Hasegawa1983}
A.~Hasegawa and M.~Wakatani, 
Plasma Edge Turbulence,
Physical Review Letters., {\bf 50} (1983), 682-686.

\bibitem{Numata2007}
R.~Numata et al., 
Bifurcation in electrostatic resistive drift wave turbulence,
Phys. Plasmas., {\bf 14} (2007), 102312.

\bibitem{Coley2022}
C.~Coley and J.~Metzler, 
Approximating Turbulent Flows using Knowledge-guided Tensor Decomposition, 
AIAA AVIATION 2022 Forum (2022).

\bibitem{Alfaro202}
C.~Alfaro-Isac, S.~Izquierdo-Estallo, and J.~Sierra-Pallares, 
Reduced-order modelling of equations of state using tensor decomposition for robust, accurate and efficient property calculation in high-pressure fluid flow simulations, 
J. Supercritical Fluids, {\bf 165} (2020), 104938.

\bibitem{Zhang2022}
G.~Zhang, et al.,
Three-dimensional wind field reconstruction using tucker decomposition with optimal sensor placement,
Energy, {\bf 260} (2022), 125098.

\end{thebibliography}
\end{document}